\documentclass[fleqn,usenatbib]{mnras}
\usepackage{amssymb}	
\usepackage{newtxtext,newtxmath}

\usepackage[T1]{fontenc}

\DeclareRobustCommand{\VAN}[3]{#2}
\let\VANthebibliography\thebibliography
\def\thebibliography{\DeclareRobustCommand{\VAN}[3]{##3}\VANthebibliography}

\usepackage{graphicx}	
\usepackage{amsmath}	
\usepackage{bm}

\title[Constraining Vainsthein]{Probing Vainsthein-screening gravity with galaxy clusters using internal kinematics and strong and weak lensing}

\author[L. Pizzuti et al.]{
Lorenzo Pizzuti,$^{1}$\thanks{E-mail: pizzuti@oavda.it}
Ippocratis D. Saltas,$^{2}$
Keiichi Umetsu$^{3}$
and Barbara Sartoris$^{4}$
\\
$^{1}$Osservatorio Astronomico della Regione Autonoma Valle d'Aosta,  Loc. Lignan 39, I-11020, Nus, Italy\\
$^2$ CEICO, Institute of Physics of the Czech Academy of Sciences, Na Slovance 2, 182 21 Praha 8, Czechia\\
$^{3}$Academia Sinica Institute of Astronomy and Astrophysics (ASIAA), No.~1, Section~4, Roosevelt Road, Taipei 10617, Taiwan\\
$^{4}$Universit{\"a}ts-Sternwarte M{\"u}nchen, Fakult{\"a}t f{\"u}r Physik, Ludwig--Maximilians Universit{\"a}t, Scheinerstrasse 1, 81679 M{\"u}nchen, Germany
}

\date{Accepted XXX. Received YYY; in original form ZZZ}

\pubyear{2021}

\begin{document}
\label{firstpage}
\pagerange{\pageref{firstpage}--\pageref{lastpage}}
\maketitle

\begin{abstract}
We use high-precision combined strong/weak lensing and kinematics measurements of the total mass profiles of the observed galaxy clusters MACS~J1206.2-0847 and Abell~S1063, to constrain the relativistic sector of the general DHOST dark energy theories, which exhibit a partial breaking of the so called Vainsthein screening mechanism, on the linear level of scalar fluctuations around a cosmological background. In particular, by using the \textsc{MG-MAMMPOSSt} framework developed in Pizzuti et al. 2021,   for the kinematics analysis of member galaxies in clusters along with lensing mass profile reconstructions  
we provide new constraints on the coupling $Y_2$ which governs the theory's relativistic contribution to the lensing potential. The new bound from  the combination of kinematics and lensing measurements of MACS 1206, $Y_2=-0.12^{+0.66}_{-0.67}$ at $2\sigma$, provides about a 2-fold improvement on previous constraints. In the case of Abell~S1063 a $>2\sigma$ tension with the GR expectation arises. We discuss this in some detail, and we investigate the possible sources of systematics which can explain the tension. We further discuss why the combination of kinematics of member galaxies with lensing is capable of providing much tighter bounds compared to kinematics or lensing alone, and we explain how the number density profile of tracers, as well as the choice of the velocity anisotropy profile affects the final results.

\end{abstract}


\section{Introduction}
Scalar field theories aiming to explain dark energy in a cosmological framework play a fundamental role towards understanding the nature of gravity at large scales and the problem of the late-time accelerated expansion of the universe (e.g. \citealt{riess98,perlmutter99}). Scalar field extensions of General Relativity (GR) possess a rich phenomenology (see e.g. \citealt{Cantata21}), and they typically predict departures from Newtonian dynamics at large scales as well as from the GR predictions on gravitational lensing, a consequence of the presence of new dynamical degrees of freedom.
In this context, galaxy clusters are excellent natural laboratories in devising observational tests for modified gravity (MG) theories for two main reasons: on the one hand, they lie in the marginal regime between cosmological and astrophysical scales, where possible departures from GR could leave detectable imprints. On the other hand, they allow for jointly constraining both the relativistic and non-relativistic sectors of the gravitational interaction, through lensing and internal kinematics (of gas or member galaxies) respectively. For this purpose, galaxy clusters have been previously used extensively to test alternative gravity scenarios through the dynamics of the intra-cluster gas and weak lensing \citep{Terukina2012, Terukina:2013eqa,Wilcox:2015kna, Wilcox:2016guw,Sakstein:2016ggl,Salzano:2017qac}, density distribution of galaxy clusters \citep{Schmidt2009, Rapetti2010, Lombriser:2010mp, Rapetti2011, Cataneo2015, Ferraro2011, Cataneo:2016iav}. 
In a series of previous papers (e.g. \citealt{Pizzuti:2016ouw,Pizzuti:2019wte}) we showed that the velocity field of member galaxies in clusters, combined with weak and strong lensing observations allows for the reconstruction of the local gravitational potentials, which in turn allow to test the assumptions about the underlying gravity theory or input physics. Such method relied on the utilisation of the newly introduced code \textsc{MG-MAMPOSSt} and was exposed in full details in \citet{Pizzuti2021}.

In the present work, we will apply this method on high-quality spectroscopic and imaging data of two massive galaxy clusters analysed in detail as part of the Cluster Lensing And Supernova survey with Hubble
 \citep[CLASH;][]{Postman2012} and the spectroscopic follow-up with the Very Large Telescope \citep[CLASH-VLT;][]{Rosati2014} programs:  MACS J1206.2-0847 (hereafter MACS 1206) at redshift $z= 0.44$ and Abell~S1063  at $z= 0.35$. We will focus on the general theories for dark energy known as Degenerate Higher-Order Scalar Tensor theories (DHOST hereafter, see e.g. \citealt{BenAchour:2016fzp}). As it is well known, this family of dark energy models predicts a modification to the linear scalar potentials governing the kinematics of member galaxies and deflection of light through gravitational lensing respectively. 
Our analysis is not able to provide a competitive bound on the fifth-force coupling $Y_1$ associated with the Newtonian potential $\Phi$, a result expected from our previous forecasting analysis presented in \cite{Pizzuti2021}. Instead, we will show that the combination of kinematics and lensing measurements provides an improvement on the complementary coupling $Y_2$, which is associated with the relativistic potential $\Psi$. The latter constraint turns out to be $\sim 6$- times tighter than the one derived using lensing observations alone and about $2$ times tighter than previous constraints in literature.
\\
Our work splits as follows: In Section \ref{sec:setup} we introduce our theoretical setup, and the relevant equations we use.
In Section \ref{sec:dataset} we briefly summarize the characteristics of MACS 1206 and   Abell~S1063.
In Sections \ref{sec:analysis} we first present the results based on the individual internal kinematics and lensing analyses, before we proceed with the combined kinematics+lensing analysis and the derivation of the relevant constraints, further discussing the impact of possible systematics in the kinematics analysis.
Finally, in Section \ref{sec:summary}, we draw our main conclusions and we elaborate on possible developments of this work.

Our main results are shown in Table \ref{tab:results} and Figures \ref{fig:resultMACS},\ref{fig:stack}, \ref{fig:resultRXJ}.

\section{Theoretical setup} \label{sec:setup}
The theories we are interested in fall within the general class of DHOST theories which predict a fifth force operating within massive sources such as a galaxy cluster. The underlying reason for this phenomenological effect is the breaking of the so--called Vainshtein screening mechanism (see e.g.  \citealt{Koyama:2015oma}), in these theories. The screening mechanism is responsible for suppressing the effect of the fifth force in the vicinity or within massive sources so that agreement with Newtonian gravity is restored. This way, solar-system tests are evaded. However, the mechanism is broken within massive sources for this family of theories. 

The fifth force, propagated by the scalar degree of freedom, affects the Poisson equations associated to the Newtonian potential $\Phi$, as well as the relativistic one, $\Psi$, according to \citep{Kobayashi:2014ida,Crisostomi:2017lbg,Dima:2017pwp},
\begin{equation}\label{eq:BH_Mamon}
\frac{\text{d} \Phi(r)}{\text{d}r} =  \frac{G M(r)}{r^2} \left[1+\frac{3}{4}Y_1\left(\frac{\rho(r)}{\bar{\rho}(r)}\right)\left(2+\frac{\text{d}\ln \rho}{\text{d}\ln r}\right)\right],
\end{equation} 

\begin{equation} \label{eq:BH_lensing}
    \frac{\text{d} \Psi(r)}{\text{d}r} =\frac{G M(r)}{r^2}\left[1-\frac{15}{4}Y_2\left(\frac{\rho(r)}{\bar{\rho}(r)}\right)\right].
\end{equation}
In the above equations, we have assumed spherical symmetry. $\bar{\rho}(r)$ is the (spatially) average density at radius $r$ from the center of the galaxy cluster, and $Y_1, Y_2$ correspond to the dimensionless fifth-force couplings. Finally, $G$ is the Newton's constant.
Whereas the dynamics of member galaxies in the cluster are governed by the potential $\Phi$, lensing is sourced by the combination 
\begin{equation} \label{eq:lenspot}
\frac{d}{dr} \Phi_{\rm{lens}} = \frac{1}{2}\frac{d}{dr}(\Phi + \Psi).
\end{equation}
Therefore, kinematical observations allow for contraints on $Y_1$, while lensing constrains both $Y_1$ and $Y_2$. The right-hand side of above equation can be expressed in terms of the density profile $\rho(r)$ according to the relevant equations for $\Phi$ and $\Psi$ above. The dominant source of pressureless matter density in the cluster comes from dark matter, which density we choose to model with a Navarro-Frenk-White (NFW) of \cite{navarro97} profile as
\begin{equation}\label{eq:NFWdens}
\rho(r)=\frac{\rho_\text{s}}{r/r_\text{s}(1+r/r_\text{s})^2},
\end{equation}
with $\rho_\text{s}$ is a characteristic density and $r_\text{s}$ the radius at which the logarithmic derivative of the density profile takes the value $-2$. The NFW profile has been shown to provide an overall good agreement with observations and simulations over a  broad range of scales in GR (e.g. \citealt{Biviano01,Umetsu16,Peirani17}) and in modified gravity (e.g. \citealt{Lomb12,Wilcox:2016guw}). Moreover, the GR analyses with lensing and internal kinematics of both clusters indicate that the total mass profile is well fitted by the NFW model (\citealt{Biviano01,Umetsu16,Caminha2017,Sartoris20}).
Under the assumption of a NFW profile, we can re-write the equation for the potential $\Phi$ in an effective way as
\begin{equation} \label{eq:massdyn}
  \frac{\text{d}\Phi}{\text{d}r} \equiv \frac{G M_{\text{dyn}}}{r^2}=\frac{G}{r^2}\left[ M_{\rm{NFW}}(r)+M_1(r)\right],   
\end{equation}
which serves as a definition of the dynamical mass $M_{\text{dyn}}$. Notice that, $G$ here is still Newton's constant as measure in the solar system. The fifth-force contribution $M_1$ is defined in terms of the NFW parameters as
\begin{equation} 
    M_1(r)= M_{200}\frac{Y_1}{4}\frac{r^2(r_\text{s}-r)}{(r_\text{s}+r)^3}\times[\ln(1+c)- c/(1+c)]^{-1}.
\end{equation}
where $c=r/r_s$ is the concentration and $M_{200}$ is the mass of a sphere of radius $r_{200}$ enclosing an average density 200 times the critical density of the universe at that redshift. In a similar fashion, the relevant expression for the lensing mass can be found by computing 
\begin{equation}
M_{\text{lens}}(r) =\frac{r^2}{2G}\left[\frac{\text{d}\Psi}{\text{d}r}+\frac{\text{d}\Phi}{\text{d}r}\right]. \label{eq:M_lens} 
\end{equation}

\begin{equation*} 
    M_{\text{lens}}=M_{\text{NFW}}+\frac{r^2M_{200}\left[Y_1(r_\text{s}-r)-5Y_2(r_\text{s}+r)\right]}{4[\log(1+c_{200})-c_{200}/(1+c_{200})]}\frac{1}{(r_\text{s}+r)^{3}},
\end{equation*}
which can be effectively re-expressed in terms of the dynamical mass as
\begin{equation}\label{eq:Mlens}
M_{\text{lens}}    \equiv M_{\text{dyn}}+M_2, 
\end{equation}
with $M_2$ the contribution from the fifth force defined through
\begin{equation}
   M_2=\frac{r^2M_{200}}{8(r_\text{s}+r)^{3}}\frac{Y_1(r-r_\text{s})-5Y_2(r_\text{s}+r)}{[\ln(1+c)-c/(1+c)]}. 
\end{equation}
In view of the above equations, it is important to emphasise again that, whereas the fifth force effect enters the dynamical mass only through the coupling $Y_1$, the lensing mass is affected by both $Y_1$ and $Y_2$. This is expected, since lensing is sourced by the combination of the two potentials $\Phi$ and $\Psi$, eq. \eqref{eq:lenspot}. 
Note also that, with gravitational lensing observations, one reconstructs the projected surface mass density profile $\Sigma(R)$, where $R$ is the projected radius from the cluster center. We refer to e.g. \citet{Umetsu20} for an explicit discussion of the physics and mathematical framework. 

Therefore, the set of free parameters to be constrained in a combined kinematics+lensing analysis are the NFW ones together with the fifth force couplings,
\begin{align}
(r_{200}, r_{\text{s}}; \, Y_1, Y_2).
\end{align}
To the above parameters, we will add the free parameter coming from the anisotropy profile in the analysis of the kinematical data, as we will explain below.

\section{Datasets} \label{sec:dataset}

In this paper, we apply the methodology described in Section \ref{sec:analysis} to the two very well studied galaxy clusters MACS 1206 and Abell~S1063 with very high-quality imaging and spectroscopic data. We describe hereafter the principal characteristics of these two clusters and the data-sets we used to perform the kinematic and the lensing analyses.

Both clusters were selected to be part  of  the  HST  treasury  program  CLASH \citep{Postman2012}. Moreover an extensive spectroscopic follow-up campaign  has been conducted within the CLASH-VLT project \citep{Rosati2014} with the VIMOS spectrograph and additional spectroscopic data have been taken with the MUSE integral field instrument in the cluster core \citep{Caminha2017}.

The analysis of \citet{girardi15} indicates that MACS 1206 is an overall relaxed systems with some minor substructures in the projected distribution of the galaxies, as further confirmed by \cite{Lemze01}. The dynamical relaxation condition is suggested also by the nearly concentric distribution of the projected mass components (as shown in e.g. \citealt{UmetsuMACS}) .
\citet{Biviano01} carried out a parametric reconstruction of the mass profile from internal kinematics analysis using a sample of 592 cluster members identified among a total of 2749 galaxies with reliable redshift measured with the VLT/VIMOS spectrograph. They perform a Maximum likelihood fit by using the \textsc{MAMPOSSt} technique (\citealt{Mamon01}, see Sec. \ref{sec:dynamics}) to determine the total mass down to the cluster core ($\sim 0.05\,\text{Mpc}$). Among the models analysed, the NFW profile provides the highest likelihood from the fit.

As for Abell~S1063 \citet{Abell89}, also known as RXJ2248.7$-$4431  (\citealt{degrandi99}), the combined data-sets from VIMOS and MUSE spectrographs consist in 3850 observed redshifts, from which a sample of 1234 cluster members were selected. 
The total spectroscopic sample of this cluster at redshift $z=0.346$ is presented in  \citet{Mercurio21} \citet[see also][]{Caminha16}.
With this hugh data set in \citet{Sartoris20}, they reconstructed the kinematic mass profile from $\sim 1$~kpc up to the virial radius. In such analysis, they disentangled the contribution of the DM from the one of the stellar  mass  profile  of  cluster  members,  from the contribution Brightest Central Galaxy,  and  from the contribution of the hot intra-cluster gas. They founded that the DM profile is well described by a generalized NFW (gNFW) model with an exponent $\gamma_{DM}=0.99\pm0.04$ close to the value of the standard NFW. Moreover they compared the mass total profile as obtained from the full dynamical analysis with the mass profiles obtained from the Chandra X-ray data and the strong+weak lensing analysis of \citet{Umetsu16} and \citet{Caminha16}. They found that the profiles are in overall good agreement, whereas a discrepancy with the (non-parametric) weak-lensing results at $~0.3 $~Mpc has been found (see Figure~7 of \citealt{Sartoris20}). 
This is the clue of the residual of a recent off-axis merger, as indicated by the analysis of \citet{Gomez12} and also suggested by the elongated shape of the X-ray emission. A detail description of the dynamical status of Abell~S1063 has been carried out by \citet{Mercurio21}  \citep[see also][]{Bonamigo18}. They pointed out as Abell~S1063 is far from being a relaxed system, with a recent off-axis merger event and a non-guassian velocity dispersion along the line of sight. As we will see in Section \ref{sec:combined}, deviation from dynamical relaxation can be the source of spurious detections of modified gravity.  

Furthermore, we employ the CLASH lensing data products presented in \citet{Umetsu16}, \citet{Caminha16}, and \citet{Caminha2017} for our gravitational lens modeling of MACS~1206, Abell~S1063, and the stacked ensemble of CLASH X-ray-selected clusters. A brief overview of the lensing data is given in Section~\ref{sec:lens}.

\section{Methodology and Results}
\label{sec:analysis}
In this section we present the methodology used for the lensing and kinematics analysis, before we proceed with the presentation of our results. From now on, we will be referring to the the description of the galaxy cluster assuming that the total (effective) mass is parametrised as a NFW profile and in the presence of the fifth force as the {\bf MG-NFW} model.

\subsection{Strong and weak lensing analysis} \label{sec:lens}

The lensing analysis of \citet{Umetsu16} is limited to a subsample of 16 X-ray-selected and 4 high-magnification CLASH clusters for which wide-field weak-lensing data from ground-based observations \citep{Umetsu14} are available. Both MACS~1206 and Abell~S1063 belong to the CLASH X-ray-selected subsample of \citet{Umetsu16}.

\citet{Umetsu16} combined wide-field weak-lensing data obtained primarily with Suprime-Cam on the 8.2~m Subaru telescope \citep{Umetsu14} and small-scale weak and strong lensing data from the 16-band \textit{Hubble Space Telescope} (HST) observations \citep{Zitrin2015}.  \citet{Umetsu14} derived weak-lensing shear and magnification constraints\footnote{See Ref. \citealt{Umetsu20} for a general review of cluster--galaxy weak lensing.} over the radial range $\theta\in [0.9, 16]$~arcmin for all clusters observed with Subaru and [$0.9,14$]~arcmin for Abell~S1063 observed with ESO/WFI. \citet{Zitrin2015} constructed detailed mass models of all CLASH clusters from a joint analysis of HST strong and weak-shear lensing data. \citet{Umetsu16} combined all these lensing constraints for individual clusters to reconstruct surface mass density profiles $\Sigma(R)$ measured in a set of cluster-centric radial bins using the cluster lensing mass inversion (\textsc{clumi}) code \citep{umetsu11,Umetsu13}. in which lensing constraints are combined a posteriori in the form of azimuthally averaged radial profiles. \citet{Umetsu16} accounted for various sources of errors in their analysis. The total covariance matrix $C$ includes four terms: (i) measurement errors, (ii) uncertainties due primarily to the residual mass-sheet uncertainty,  (iii) cosmic noise due to projected large-scale structure uncorrelated with the cluster, and (iv) statistical fluctuations of the cluster lensing signal due to halo triaxiality and correlated substructures. 

\citet{Umetsu16} found that the stacked lensing signal of the CLASH X-ray-selected subsample is well described by a family of cuspy, outward-steepening density profiles, such as the NFW and Einasto models. Of these, the NFW model best describes the CLASH lensing data. 

Throughout our statistical analysis, we use the present density parameters of $\Omega_\mathrm{m}=0.3$ in matter and $\Omega_\Lambda=0.7$ in the cosmological constant $\Lambda$. For the fitting radial range we limit ourselves to $R\le 2h^{-1}$~Mpc \citep[see][]{Umetsu14,Umetsu16}. For parameter inference of the MG-NFW model, we employ uniform priors of $\log_{10}(M_{200}/h^{-1}M_\odot)\in [14,16], \log_{10}c_{200}\in [0, 1], Y_1\in [-0.67, 8], Y_2\in [-8,8]$. We notice here that, although 
the coupling $Y_1$ is constrained down to $\sim 0.1$ precision in the previous literature (see Section \ref{sec:dynamics}), we choose to set our relevant priors wide enough in order to fully understand the capabilities of our method, i.e the combination of kinematics+lensing data, to constrain alternative gravity.

We use a Markov chain Monte Carlo (MCMC) approach
with Metropolis$-$Hastings sampling to obtain well-characterized inference of the model parameters, given the data and the priors stated above. An MG-NFW posterior chain of length $10^6$ was produced for each cluster from MCMC sampling. Moreover, we consider a stacked ensemble of 15 CLASH X-ray-selected clusters excluding MACS~1206 (but including Abell~S1063) at a weighted mean redshift of $\langle z_l\rangle=0.32$. We constrain the coupling parameters ($Y_1, Y_2$) and their effective NFW parameters ($r_{200}, r_\mathrm{s}$) from their stacked lensing profile $\langle \Sigma(R)\rangle$ (for the stacking procedure, see \citealt{Umetsu16}).  In this work, we will perform a joint analysis of MACS~1206 in combination with the stacked lensing posterior constraints on the coupling parameters ($Y_1,Y_2$). 

\begin{figure*}
 \begin{center}
  \includegraphics[width=0.32\textwidth,clip]{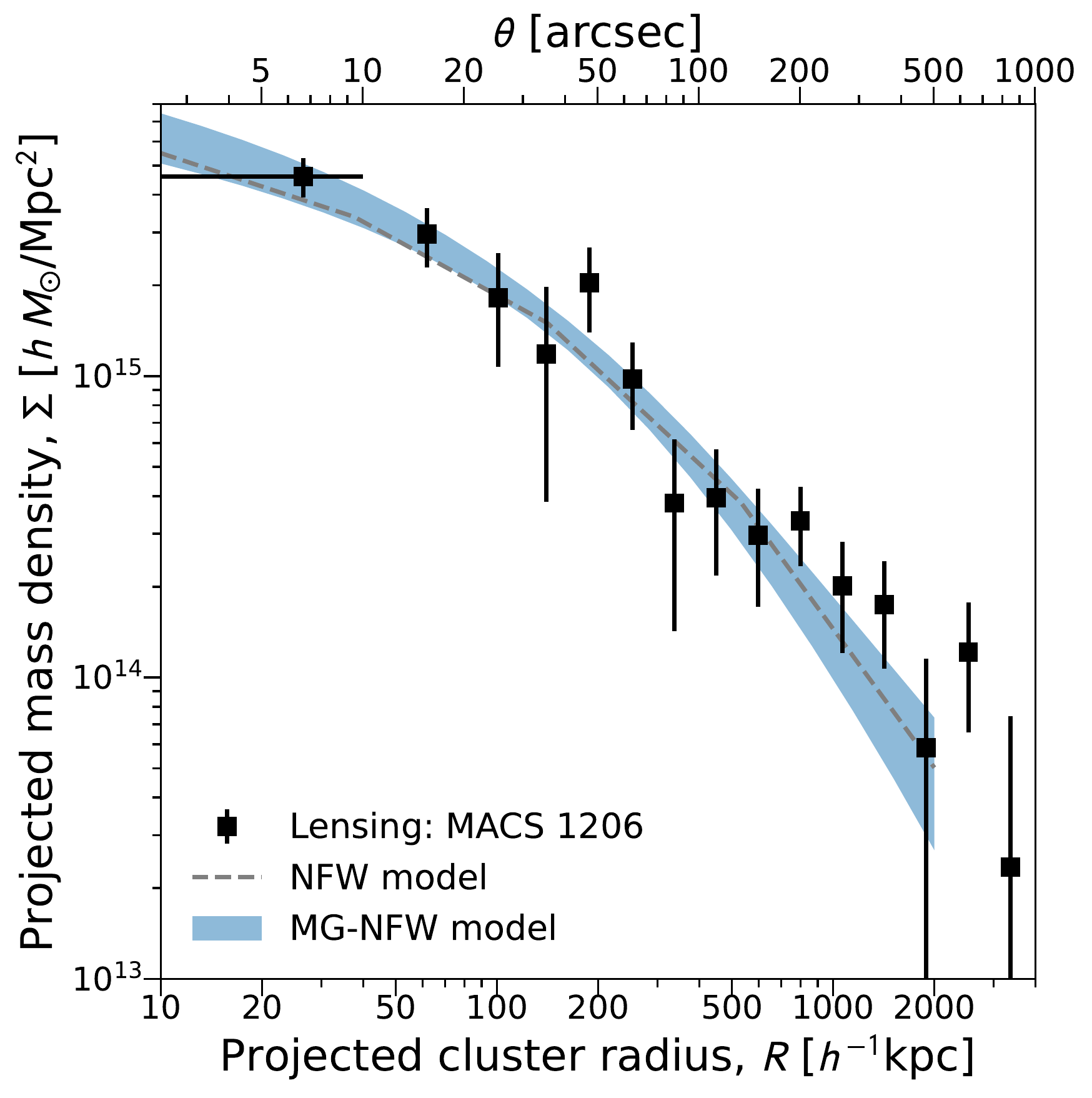}
  \includegraphics[width=0.32\textwidth,clip]{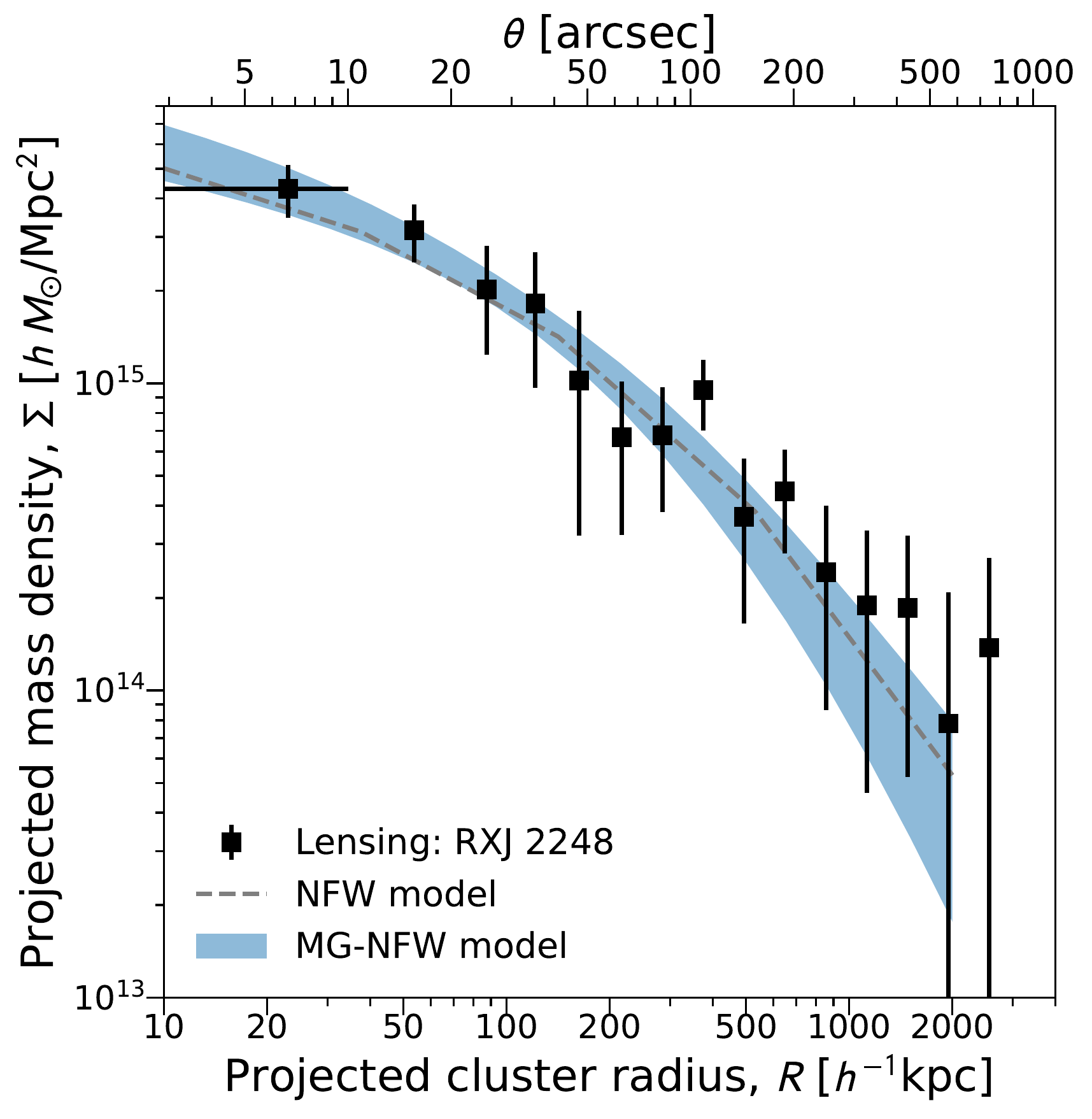}
  \includegraphics[width=0.32\textwidth,clip]{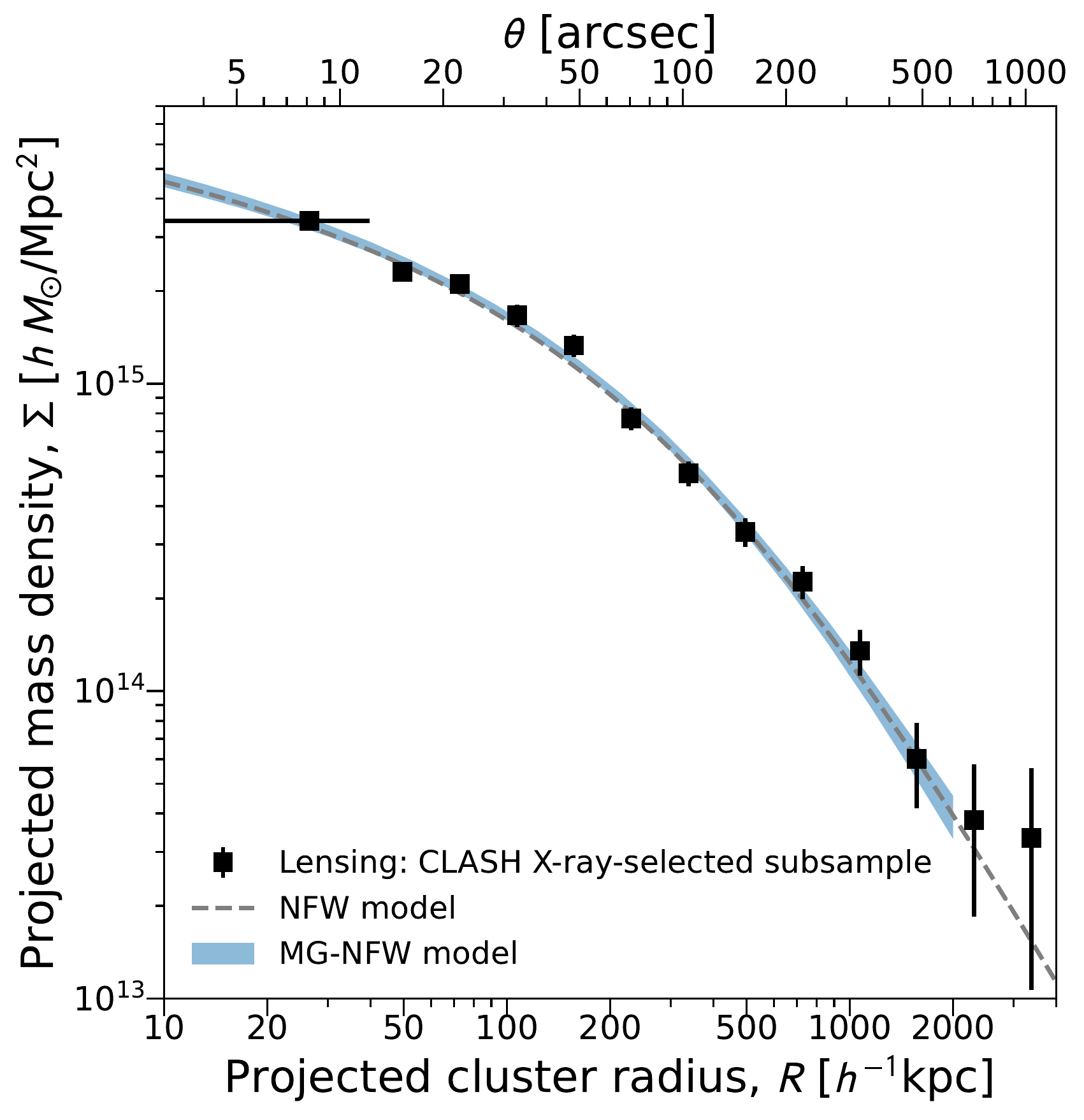}
 \end{center}
\caption{\label{fig:lens}
Projected mass density profiles $\Sigma(R)$ of MACS~1206 (left panel), Abell~S1063(middle panel), and the stacked ensemble of 15 CLASH X-ray-selected clusters (right panel). Black squares with error bars in each panel represent the binned $\Sigma(R)$ profile reconstructed from a joint analysis of CLASH weak and strong lensing data sets \citep{Umetsu16}.  In each panel, the central bin $\Sigma(<R_\mathrm{min})$ inside the minimum measurement radius $R_\mathrm{min}$ is marked with a horizontal bar.  The error bars represent the $1\sigma$ uncertainty from the diagonal part of the total covariance matrix $C$.  The gray dashed line in each panel shows the posterior mean profile of the NFW model. The blue shaded area in each panel shows the marginalized $1\sigma$ confidence region of the MG-NFW model.}
\end{figure*}

Figure~\ref{fig:lens} shows the binned projected mass density profiles $\Sigma(R)$ (black squares) of MACS~1206, Abell~S1063, and the stacked ensemble of 15 CLASH X-ray-selected clusters derived from a joint analysis of CLASH weak and strong lensing data sets \citep{Umetsu16}. For each cluster, we show the marginalized $1\sigma$ confidence region of the MG-NFW model (blue shaded area) derived from the posterior chain. For comparison, we also plot the mean posterior profile of the NFW model (dashed line).

\begin{figure}
\centering
\includegraphics[width=\columnwidth]{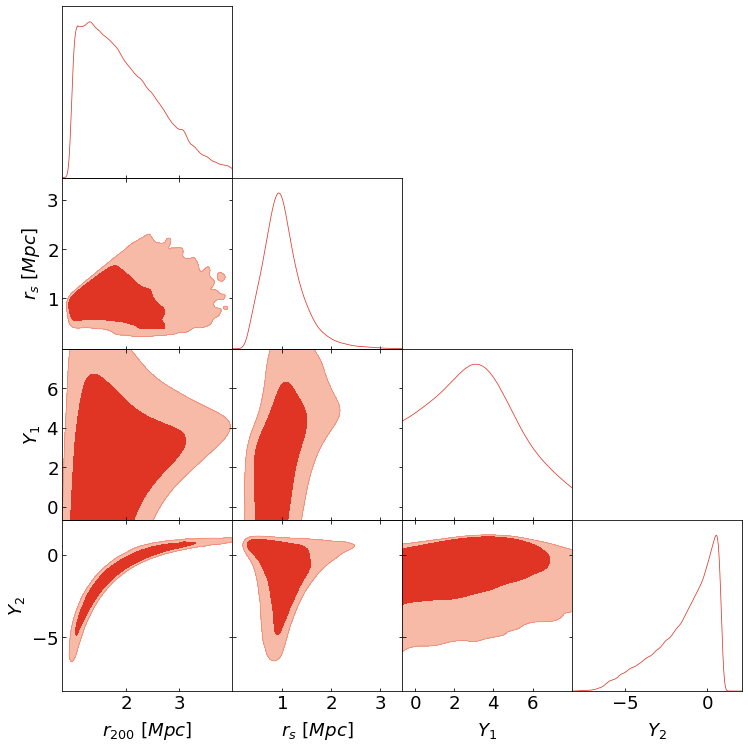}
\caption{\label{fig:MACSlens} Marginalized posteriors of the MG-NFW parameters in Vainsthein gravity from the strong+weak lensing analysis of the cluster MACS 1206. Darker and lighter filled areas refer to $1\sigma$ and $2\sigma$ confidence regions respectively. To facilitate the comparison with the internal kinematics analysis, the distributions of $\log_{10}(M_{200}/h^{-1}M_\odot)$ and $\log_{10}c_{200}$ are rephrased in terms of $r_{200}$ and $r_\text{s}$. }

\end{figure}

\begin{figure}
\centering
\includegraphics[width=\columnwidth]{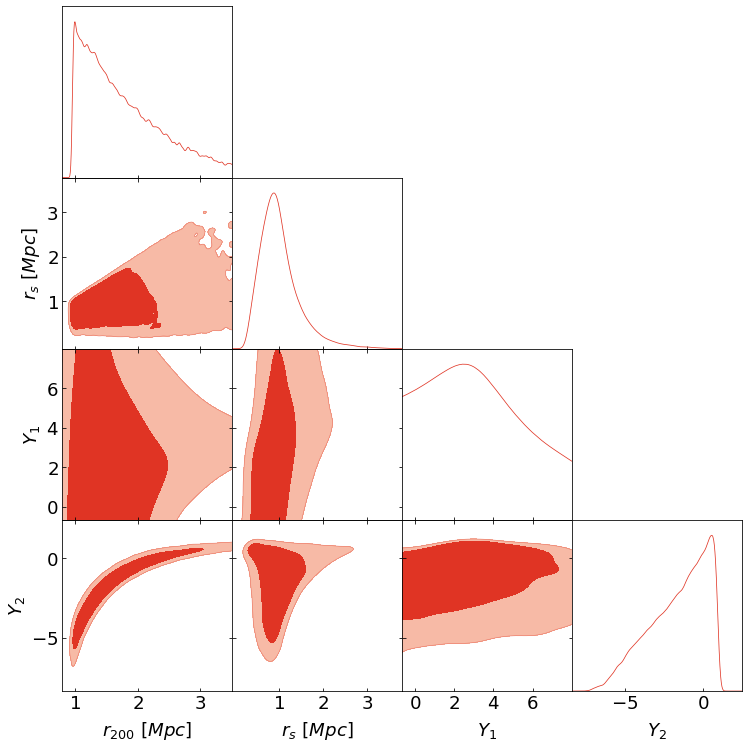}
\caption{\label{fig:RXJlens} Marginalized posteriors of the parameters $(r_{200},\,r_\text{s},\,Y_1,\,Y_2)$ in Vainsthein gravity from the strong+weak lensing analysis of the cluster Abell~S1063. Contours are color-coded according to the same prescription as in Figure \ref{fig:MACSlens}.}

\end{figure}

The marginalized distributions of the MG-NFW parameters are shown in Figures~\ref{fig:MACSlens} and \ref{fig:RXJlens} for MACS 1206 and Abell~S1063, respectively. The MG couplings $Y_1$ and $Y_2$ exhibit a strong degeneracy with the mass profile parameters, which prevent from placing any stringent bound on $Y_1$ for both clusters. In the case of $Y_2$, we can still derive a $2\sigma$-constraint of 
\begin{equation}
Y_2=-1.3^{+2.3}_{-3.6} \; (\rm{MACS}\;\rm{1206}) , \hspace{0.2cm}  Y_2=-1.6^{+2.6}_{-3.5} \; (\rm{Abell}\;\rm{S1063}), 
\end{equation}
while for the virial radius and the scale radius we obtain:
\begin{equation*}
r_{200}=2.0^{+1.47}_{-1.0}\,\text{Mpc},\, \hspace{0.2cm} r_\text{s}=1.03^{+0.83}_{-0.74}\,\text{Mpc} \; (\rm{MACS}\;\rm{1206}) ,
\end{equation*}
\begin{equation}
r_{200}=1.85^{+1.7}_{-0.91}\,\text{Mpc},\,  \hspace{0.2cm} r_\text{s}=1.01^{+0.94}_{-0.75}\, \text{Mpc} \; (\rm{Abell}\;\rm{S1063}), 
\end{equation}
where the central values have been chosen as the median of the distributions. Note that the long negative tail in the marginalized posteriors of $Y_2$ is mainly due to the degeneracy with $r_{200}$, which can be explained by a direct inspection of eq. \eqref{eq:Mlens}. A smaller-size cluster (i.e. a smaller $r_{200}$) can be compensated by increasing the $M_2$ term in the effective lensing mass through a negative $Y_2$, corresponding to an overall strengthening of the gravitational interaction. 

\begin{figure}
\centering
\includegraphics[width=\columnwidth]{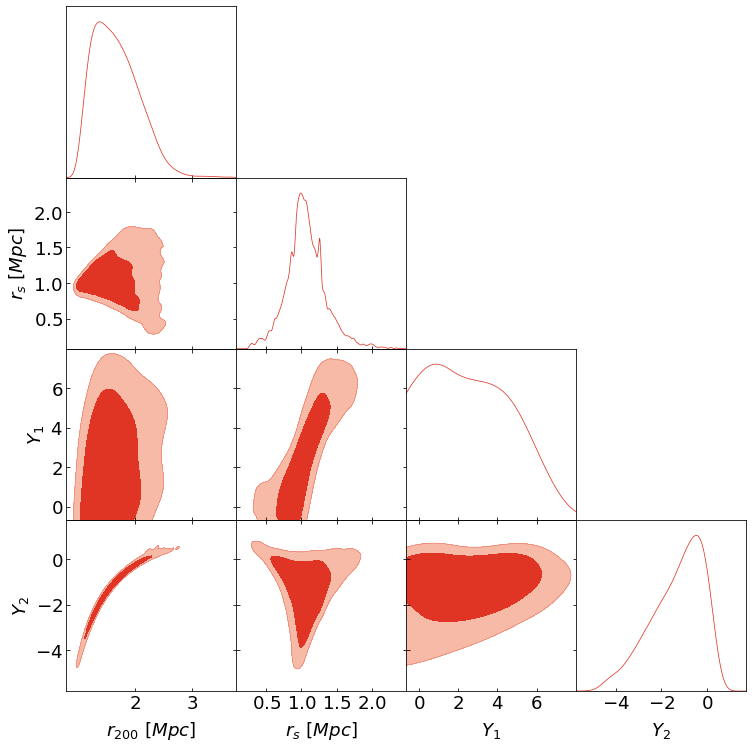}
\caption{\label{fig:lensingstack} Marginalized posteriors of the parameters $(r_{200},\,r_\text{s},\,Y_1,\,Y_2)$ in Vainsthein gravity from the stacked strong+weak lensing analysis of 15 CLAXH X-ray-selected clusters. Contours are color-coded according to the same prescription as in Figure \ref{fig:MACSlens}.}

\end{figure}

In Figure~\ref{fig:lensingstack} we show the marginalized distribution obtained from the stacked weak+strong lensing analysis of 15 CLASH X-ray-selected clusters (right panel of Figure~\ref{fig:lens}). In this case there is a slight improvement on the constraints for both couplings:
\begin{equation*}
    Y_1<6.31\,,\;\hspace{0.2cm} Y_2=-1.3^{+1.9}_{-2.4}.  
\end{equation*}
which is however too modest to place competitive bounds on these parameters. 


\subsection{Galaxy kinematics analysis: \textsc{MG-MAMPOSSt}}
\label{sec:dynamics}
The kinematics analysis is based on utilisation of our recently released code \textsc{MG-MAMPOSSt}, a version of the \textsc{MAMPOSSt} method of \citet{Mamon01} which reconstruct the cluster mass profiles in modified gravity by using the kinematics information of member galaxies. The code has been presented in detail in our previous work \citet{Pizzuti2021}. We refer to the latter paper for details on the framework and numerical method used, and here we restrict ourselves only to a brief overview of the approach. 

\textsc{MG-MAMPOSSt} relies on a Maximum Likelihood technique to fit data extracted from the projected phase space $(R,v_z)$ of the cluster's member galaxies, tracers of the total gravitational potential. Above, $R$ is the projected position of the galaxies with respect to the cluster center and $v_z$ the velocity measured along the line-of-sight (l.o.s.) in the rest frame of the cluster. 
Assuming dynamical relaxation and spherical symmetry, the code solves the (time-independent) Jeans equation for a given model of the gravitational potential $\Phi$:
\begin{equation}\label{eq:ippo-jeans}
\frac{\text{d} (\nu(r)\sigma_r^2)}{\text{d} r}+2\beta(r)\frac{\nu\sigma^2_r}{r}=-\nu(r)\frac{\text{d} \Phi}{\text{d} r},
\end{equation}
to obtain the expression of the radial velocity dispersion $\sigma^2_r$.
In the above equation, $\nu(r)$ is the number density distribution of the galaxies and $\beta=(\sigma^2_\theta+\sigma^2_\phi)/2\sigma^2_r$ is the velocity anisotropy profile which accounts for the difference in the velocity dispersion components $\sigma^2_r,\,\sigma^2_\theta,\,\sigma^2_\phi$, along the radial, azimuthal and tangential direction respectively. In spherical symmetry $\sigma^2_\theta=\sigma^2_\phi$, thus the expression for the anisotropy simplifies to  $\beta(r)=1-\sigma^2_\theta/\sigma^2_r$. 

In \textsc{MG-MAMPOSSt} we implement a parametric reconstruction of the velocity anisotropy profile, which is fitted together with the gravitational potential (see \citealt{Pizzuti2021} and references therein for more details). We further determine the systematic effect induced on our final results - i.e. on the joint lensing+kinematic analysis - by different choices of the parametrisation, as done in \citet{Pizzuti:2017diz}, in Section \ref{sec:combined}. 

For each cluster we consider four different models of the velocity anisotropy profile, namely:\\
the \textbf{constant anisotropy model $\text{''C''}$} 
\begin{equation}\label{eq:betac}
\beta(r)=\beta_\text{C},
\end{equation}
the \textbf{Mamon\&Lokas model $\text{''ML''}$} of \citet{MamLok05}, which provides an adequate fit to the average anisotropy profile of cluster-size halos in cosmological simulations (e.g. \citealt{mamon10})
\begin{equation}\label{eq:betaml}
\beta_\text{ML}(r)=\frac{1}{2}\frac{r}{r+r_{\beta}},
\end{equation}
where $r_{\beta}$ is a characteristic scale radius; \\
the \textbf{Tiret model  $\text{''T''}$} of \citet{Tiret2007}
\begin{equation}\label{eq:betat}
\beta_\text{T}(r)=\beta_{\infty}\frac{r}{r+r_\beta},
\end{equation}
a generalized version of the  $\text{''ML''}$ which tends to $\beta_{\infty}$ at large radii;\\
the \textbf{Opposite model  $\text{''O''}$}
\begin{equation}\label{eq:betao}
\beta_\text{O}(r)=\beta_{\infty}\frac{r-r_\beta}{r+r_\beta},
\end{equation}
of \citet{Biviano01}, which allows for negative anisotropy (i.e. tangential orbits) in the innermost region. 
For the  $\text{''T''}$ and  $\text{''O''}$ profiles, we consider the normalization $\beta_{\infty}$ as the free parameter, while the scale radius is set to be equal to $r_\text{s}$. Moreover, except for the $\text{''ML''}$ model, we will work with the scaled quantity:
\begin{equation*}
 \mathcal{A}_\infty \equiv (1-\beta_{\infty/C})^{-1/2},   
\end{equation*}

 In principle, the number density profile can be also fitted within the \textsc{MG-MAMPOSSt} procedure; however, the distribution of the tracers can be obtained by external analyses of the projected phase space, taking into consideration the completeness of the sample. In particular, we perform a Maximum Likelihood fit to the numerical distribution of the member galaxies which doesn't require a binning of the data (see e.g. \citealt{Biviano01}) assuming a projected NFW model; we then fix the value of $r_\nu$ in \textsc{MG-MAMPOSSt} as the best fit. We will discuss the effect induced by a variation of $r_\nu$ within its $1\sigma$ uncertainties for both clusters at the end of Section \ref{sec:combined}. Note that the normalization constant of $\nu(r)$ is not relevant in \textsc{MG-MAMPOSSt} as it factors out in the solution of the Jeans' equation. We obtain $r_\nu=0.89^{+0.17}_{-0.13}$ for MACS 1206 and $r_\nu=0.76^{+0.08}_{-0.07}$ (see \citealt{Sartoris20}).
 
Using the obtained expression for $\sigma^2_r$, the code computes the probability $q(R_i,v_{z(i)}| \bf{\theta})$ to observe a galaxy at projected position $R_i$ with a l.o.s. velocity $v_{z(i)}$, given the vector of parameters ${\bold{\theta}}$ which describes the chosen model(s). In the case of Vainsthein screening, ${\bf\theta}=(r_{200},r_\text{s},\beta,Y_1)$.
The final output is the likelihood, computed by combining the probability of each single galaxy as:
\begin{equation}
-\ln \mathcal{L}_{\text{dyn}}=-\sum_{i=1}^N \ln q(R_i,v_{z(i)}| \bf{\theta}).
\end{equation}
Here, $N$ is the number of member galaxies.
We apply the \textsc{MG-MAMPOSSt} to the data-sets of MACS 1206 and Abell~S1063 to fit together the mass profile parameters, the anisotropy parameter and the copuling $Y_1$ assuming that the total effective mass profile is described by the MG-NFW of eq. \eqref{eq:BH_Mamon}. We point out again that in the internal kinematics analysis no information can be obtained on $Y_2$, which appear only in the relativistic potential eq. \eqref{eq:BH_lensing}. \\
For the case of MACS 1206, we consider 375 galaxies in the (projected) radial range $[0.05\,\text{Mpc},\,1.96\,\text{Mpc}]$ (see e.g. \citealt{Biviano01,Pizzuti:2017diz}), while for Abell~S1063 the sample consists in 781 tracers within $[0.05\,\text{Mpc},\,2.36\,\text{Mpc}]$ (\citealt{Sartoris20}). In both cases, the lower bounds have been chosen to avoid the contribution of the BCG, which dominates the internal dynamics at small scales, while the upper limit ensures that the analysis is performed within the virial region of the cluster where the Jeans equation is valid. We checked that changing these values within a $\sim 7\%$ range does not affect the final results.

We perform an MCMC sampling of the parameter space assuming uniform priors in the range of each parameter as $r_{200}\in [1.5\,\text{Mpc},4.5\,\text{Mpc}],\, r_\text{s} \in [0.05\,\text{Mpc}, 4.0\,\text{Mpc}],\, \mathcal{A}_\infty \in [0.5,7.1],\,Y_1 \in [-0.67,8.0]$. We consider the same bounds for both clusters checking that varying the ranges does not alter the final posterior. As for the velocity anisotropy parameter, we adopt slightly different priors in the case of the "ML" model, where the free parameter is the scale radius $r_\beta \in [0.05,4.0]$. 
 

%
%
%

\begin{figure}
\centering
\includegraphics[width=\columnwidth]{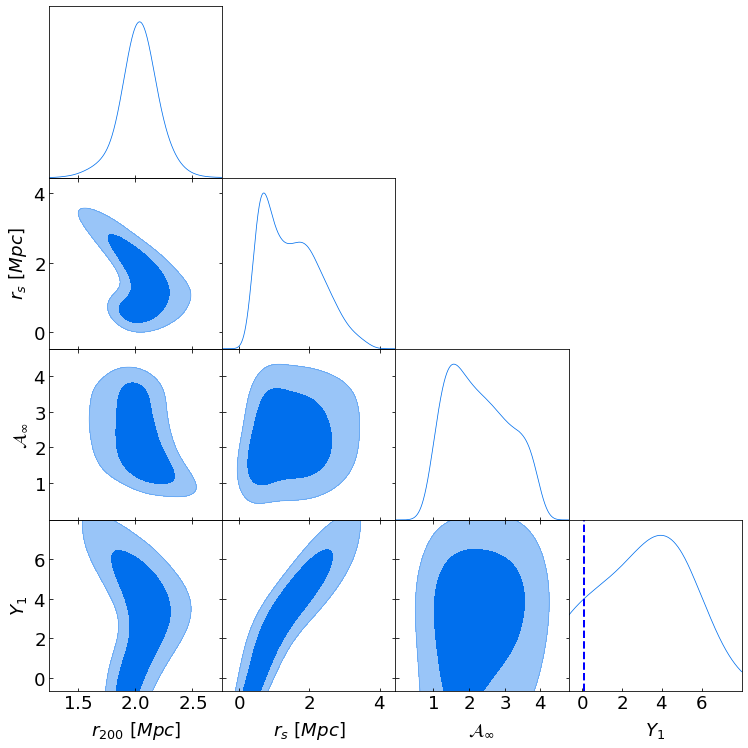}
\caption{\label{fig:MACSdyn} Marginalized distributions of the free parameters in Vainsthein gravity from the \textsc{MG-MAMPOSSt} analysis alone (i.e. from kinematics of members galaxy only) of the cluster MACS 1206, employing the "T" anisotropy model. The latter model is chosen due on the grounds of smaller value of $\chi^2$ compared to other anisotropy models. Darker and lighter filled areas refer to $1\sigma$ and $2\sigma$ confidence regions respectively. We remind here that, the fifth force coupling $Y_1$ governs the Newtonian potential, and therefore, the kinematics of the cluster. The resulting constraints on $Y_1$ are weak, and do not compete with the current ones from stellar physics which are $\sim \mathcal{O}(0.1)$. }

\end{figure}
\begin{figure}
\centering
\includegraphics[width=\columnwidth]{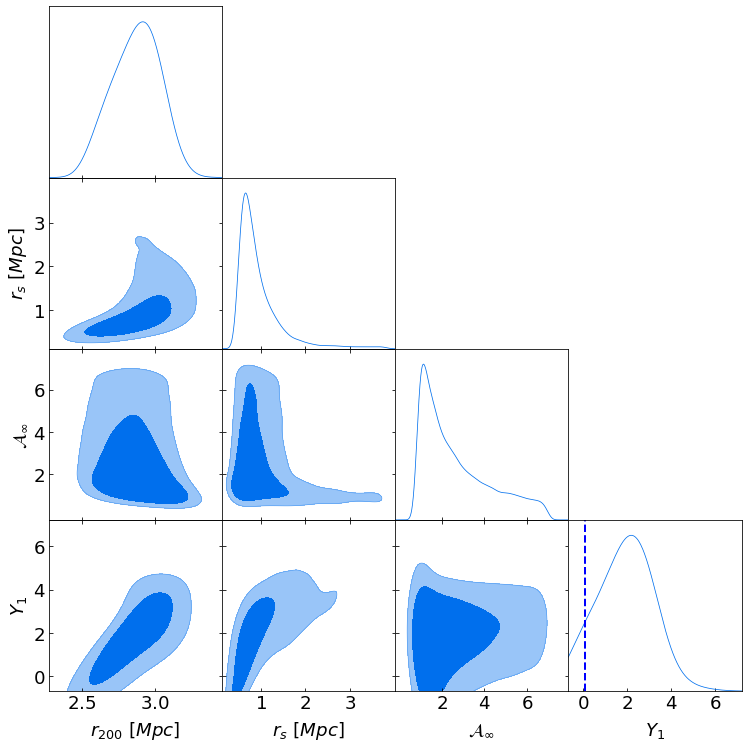}
\caption{\label{fig:RXJdyn} Marginalized distributions of the free parameters in Vainshtein gravity from the analysis of the internal kinematics for the cluster Abell~S1063 with "O" anisotropy model. The latter model is chosen for similar reasons to the choice for MACS 1206, i.e due to the yield of a better $\chi^2$ compared to other models. Darker and lighter filled areas refer to $1\sigma$ and $2\sigma$ confidence regions respectively. The constraints on $Y_1$ for this cluster are improved compared to those of MACS 1206 (Figure \ref{fig:MACSdyn}), though still not competitive. }

\end{figure}

In both lensing and internal kinematics analysis, the condition $Y_1\gtrsim-0.6$ is set to fulfill the theoretical stability constraints for astrophysical bodies (e.g. \citealt{Babichev:2016jom}). The marginalized posteriors are shown in Figure \ref{fig:MACSdyn} and Figure \ref{fig:RXJdyn} for MACS 1206 and Abell~S1063 respectively. The plots refer to the choice of the velocity anisotropy model minimizing the $\chi^2$, which is the "T" profile for MACS 1206 and the "O" model for Abell~S1063. We point out that all the four anisotropy profiles provide similar $\chi^2$ and they do not considerably change the overall results. Here we will only discuss the outcome for the best fit models, while we will address the impact of the choice of parametrisation in the joint lensing+internal kinematic analysis, Sec. \ref{sec:combined}.

As already discussed in \citet{Pizzuti2021}, the coupling $Y_1$ is strongly degenerate with the mass profile parameters $r_\text{s}$ and $r_{200}$. This is due to the structure of the effective mass, equation \eqref{eq:massdyn}. In particular, a larger value for $Y_1$ (i.e a weakening gravity) tends to decrease the gradient of the Newtonian gravitational potential for $r>r\text{s}$, an effect which is equivalent to increasing the NFW parameters $r_{200}$ or $r_\text{s}$ at standard gravity. 

We can extract a $2\sigma$ upper limit on the coupling $Y_1$ for each of the clusters considered as
\begin{equation}
Y_1< 6.8 \; (\rm{MACS}\;\rm{1206}) , \hspace{0.2cm}  Y_1<4.3 \; (\rm{Abell}\;\rm{S1063}) . 
\end{equation}
The above bounds are at least one order of magnitude larger than the constraints obtained from stellar probes \citet{Jain:2015edg, Babichev:2016jom,Saltas:2018mxc,Sakstein:2015zoa,Sakstein:2015aac, SaksteinOscil,Sakstein:2013pda,Sakstein:2018fwz} (see also \cite{Koyama, Sakstein:2015aqx}), \footnote{For strong bounds from gravitational waves under different assumptions, see \cite{Dima:2017pwp, Creminelli:2018xsv}.} and than the recent constraints of \citet{Haridasu21} from pressure and temperature profiles of gas in 12 galaxy clusters.

We highlight approximately these constraints as the blue vertical dashed lines in the right bottom plots of Figure \ref{fig:MACSdyn} and Figure \ref{fig:RXJdyn}. Moreover, the distribution for Abell~S1063 exhibits a slight ($\sim 1 \sigma$) tension with the GR expectation $Y_1=0$. As for the standard NFW parameters and the velocity anisotropy we get:
\begin{displaymath}
r_{200}=2.04^{+0.35}_{-0.39} \,\text{Mpc}\;  \hspace{0.2cm}  r_\text{s}=1.5^{+1.5}_{-1.2}\,\text{Mpc} \; \hspace{0.2cm}  
\mathcal{A}_\infty=2.3^{+1.6}_{-1.4},
\end{displaymath}
for MACS 1206 and
\begin{equation*}
  r_{200}=2.87^{+0.31}_{-0.34}\,\text{Mpc}\; \hspace{0.2cm}  r_\text{s}=0.96^{+1.1}_{-0.63}\,\text{Mpc}\; \hspace{0.2cm}  \mathcal{A}_\infty=2.7^{+3.5}_{-2.0}, 
\end{equation*}
for Abell~S1063, where, as before, the uncertainties are at $2\sigma$ and the fiducial value is given the median of the distribution.

The degeneracy between $r_\text{s},\, r_{200}$ and $Y_1$ can be broken by combining the \textsc{MG-MAMPOSSt} posterior with the information provided by the strong+weak lensing analysis, as shown in the next Section. As we will see, although the constraints on the fifth force coupling $Y_1$ do not improve, we will be able to place competitive constraints on the coupling $Y_2$.

\subsection{Combined kinematics and lensing analysis}
\label{sec:combined}
In order to constrain the coupling parameters $Y_1$ and $Y_2$, we perform a sampling of the full likelihood,
\begin{equation} \label{eq:liketotal}
\ln \mathcal{L}_{\text{tot}}=\ln \mathcal{L}_{\text{dyn}}(r_{200},r_\text{s},\mathcal{A}_\infty, Y_1)+ \ln \mathcal{L}_\text{lens}(r_{200},r_\text{s},Y_1, Y_2),
\end{equation}
over the parameter space $(r_{200},r_\text{s},\mathcal{A}_\infty,Y_1, Y_2)$. We remind here that, the \textsc{MG-MAMPOSSt} likelihood depends only on the modified gravity parameter $Y_1$, whereas the lensing on both $Y_1, Y_2$, as dictated by the structure of the equations \eqref{eq:BH_Mamon} and \eqref{eq:BH_lensing}.  The sampling is carried out for both clusters over the $10^6$-elements lensing chains, i.e. we use the lensing posteriors as priors for the combined analysis. In particular, for each set $(r_{200},r_\text{s},Y_1, Y_2)$ we randomly choose a value for the free parameter of the velocity anisotropy profile $\mathcal{A}_\infty$, assuming the same priors as discussed in Section \ref{sec:dynamics}, and we compute the total likelihood according to eq. \eqref{eq:liketotal}.

\begin{figure}
\centering
\includegraphics[width=\columnwidth]{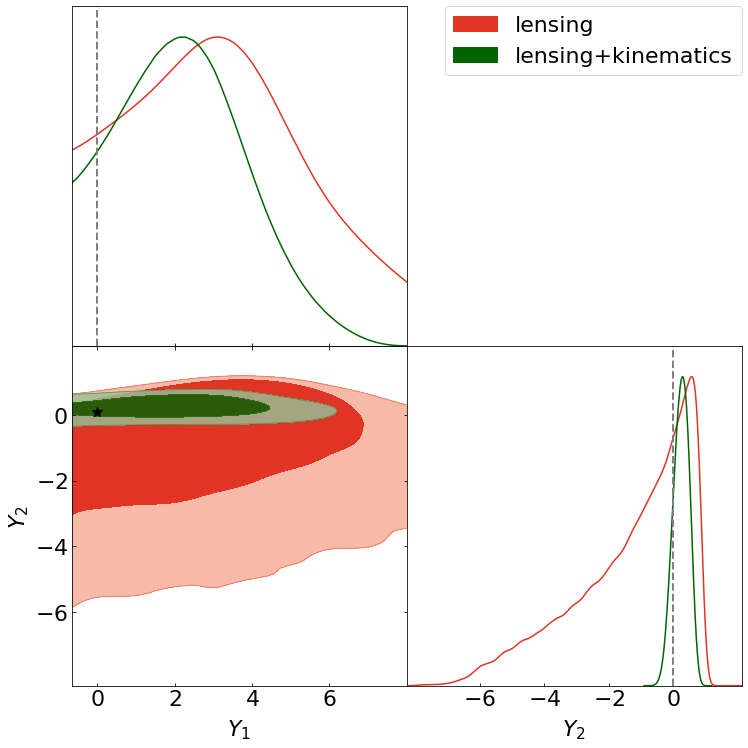}
\caption{\label{fig:resultMACS} Marginalized distributions of the two coupling constant $Y_1$ and $Y_2$ obtained from the lensing analysis (red) and from the joint lensing+kinematic analysis (green) of MACS 1206 with a ''T" anisotropy model, the latter model chosen on the same grounds as in the kinematics analysis. The inner darker regions and the outer lighter areas refer to $1\, \sigma$ and $2\, \sigma$ regions respectively. The black star and the vertical black dashed lines indicate the GR expectation values, $Y_1=0,\,Y_2=0$. The kinematics+lensing constraints on the coupling $Y_2$, associated with the linear relativistic scalar potential $\Psi$ (see eq. \eqref{eq:BH_lensing}) are competitive, and provide a {\bf 2-fold improvement} of previous constraints of this parameter. It is important to notice the strong improvement of the constraints based on kinematics+lensing compared to lensing alone -- This highlights the power of the combination of the two probes. The constraints are summarised in Table \ref{tab:results}.}

\end{figure}

As explained in \citet{Pizzuti2021}, we further exclude in the sampling all the combinations of $Y_1,Y_2$ giving rise to a negative value of the effective mass profile.
In Figure \ref{fig:resultMACS} and Figure \ref{fig:resultRXJ} we show the two-dimensional marginalized distributions of $Y_1,Y_2$ (green shaded areas, green curves) from the joint lensing+kinematic analysis compared with the constraints obtainable by lensing alone (red shaded areas, red curves) for MACS 1206 and Abell~S1063, respectively. As for the velocity anisotropy profile, we consider again as our reference case the "T" model eq. \eqref{eq:betat} for MACS 1206 and the "O" model eq. \eqref{eq:betao}  for Abell~S1063, which provide the lowest chi-square also in the joint lensing+kinematics analysis.
The $2\sigma$ constraints of both cluster for the reference case are further listed as the bold lines in Table \ref{tab:results}.

\begin{table*}
\begin{center}
\begin{tabular}{|cc|cccccc|}
   \hline
   {Anis.}& $r_{\nu}$&$r_{200}$&$r_{s}$&$\mathcal{A}_\infty$/$r_\beta$&$Y_1$&$ Y_2$ & $\Delta\chi^2$\\
    {model}      & [Mpc]  & [Mpc]  & [Mpc]&  & & & \\
    \hline
   \hline
    &&&&{MACS 1206}&&&\\
    \hline
    \hline 
 $\bm{\text{T}}$&$\bm{0.89}$&$\bm{2.14^{+0.38}_{-0.35} }$&$\bm{0.82^{+0.82}_{-0.57} }$&$\bm{2.1^{+1.7}_{-1.3} }$&$\bm{<4.85}$&$\bm{-0.12^{+0.66}_{-0.67}}$ & $\bm{0.0}$\\
   \hline
    $\text{ML}$&$0.89$&$2.19^{+0.32}_{-0.30}$&$0.75^{+0.82}_{-0.54}$&$1.3^{+2.5}_{-1.2}$&$<4.65$ & $0.02^{+0.54}_{-0.57} $ & $0.46$\\
   \hline
    $\text{O}$&$0.89$&$2.24^{+0.37}_{-0.34}$&$0.71^{+0.81}_{-0.51}$&$1.9^{+1.8}_{-1.1} $&$< 5.12 $ & $0.10^{+0.46}_{-0.51}$ & $0.81$\\
   \hline
   $\text{C}$&0.89&$2.25^{+0.40}_{-0.37}$&$0.73^{+0.82}_{-0.54}$&$1.12^{+0.40}_{-0.37} $&$< 4.61$& $0.09^{+0.58}_{-0.64}$ & $1.76$\\
    \hline
   $\text{T}$&1.06&$2.09^{+0.38}_{-0.35}$&$0.83^{+0.79}_{-0.58}$&$2.0^{+1.7}_{-1.3} $&$< 4.78$& $-0.18^{+0.69}_{-0.65}$ & $0.26$\\
   \hline
   $\text{T}$&0.76&$2.15^{+0.37}_{-0.35}$&$0.81^{+0.83}_{-0.57}$&$2.2^{+1.7}_{-1.4} $&$< 4.88$& $-0.096^{+0.58}_{-0.66}$ & $-0.02$\\
   
   \hline
   \hline
  &&&&{Abell~S1063}&&\\
    \hline
    \hline 
     $\bm{\text{O}}$&$\bm{0.76}$&$\bm{2.91^{+0.42}_{-0.43}}$&$\bm{0.99^{+1.1}_{-0.69}}$&$\bm{2.5^{+3.2}_{-2.0}}$&$\bm{2.4^{+2.5}_{-2.8}}$&$\bm{0.48^{+0.34}_{-0.34}}$ & $\bm{0.0}$\\
   \hline
    $\text{T}$&$0.76$&$2.83^{+0.40}_{-0.37}$&$1.24^{+1.2}_{-0.87}$&$2.6^{+3.5}_{-2.1}$&$2.2^{+2.5}_{-2.8}$&$0.34^{+0.43}_{-0.46} $ & $0.24$\\
   \hline
    $\text{ML}$&$0.76$&$2.91^{+0.36}_{-0.34}$&$1.10^{+1.3}_{-0.83}$&$2.4^{+3.8}_{-2.3}$&$2.2^{+2.4}_{-2.7}$&$0.45^{+0.37}_{-0.41}$ & $0.06$\\
   \hline
$\text{C}$&$0.76$&$3.00^{+0.41}_{-0.39}$&$1.12^{+1.3}_{-0.84}$&$1.06^{+0.30}_{-0.29}$&$2.3^{+2.3}_{-2.6}$&$0.49^{+0.39}_{-0.42}$ & $1.02$\\
   \hline
     $\text{O}$ & 0.68 & $2.92^{+0.40}_{-0.41}$ &$ 0.96^{+1.1}_{-0.69}$&$3.1^{+3.3}_{-2.5}$&$2.5^{+2.5}_{-2.8}$& $0.50^{+0.35}_{-0.34}$ &$-0.20$\\
   \hline
  $\text{O}$ & 0.84 & $2.90^{+0.40}_{-0.41} $ &$ 0.99^{+1.1}_{-0.71}$&$2.4^{+2.9}_{-1.8}$&$2.4^{+2.5}_{-2.8}$& $0.47^{+0.35}_{-0.34}$ &$0.17$\\
   \hline
   \hline
\end{tabular}
\caption[Results on the free parameters for the cluster MACS 1206 and Abell~S1063]{\label{tab:results}Results on the free parameters of the joint lensing+kinematics analysis for the cluster MACS 1206 (first 6 rows) and Abell~S1063 (last 6 rows). The bold characters indicate the reference models adopted  for each of the two clusters. As explained in the text, $Y_1$ is related to the fifth force coupling governing the kinematics of galaxies in the cluster, whereas a combination of $Y_1, Y_2$ controls the effect of the fifth force on the lensing. As it turns out, the combined kinematics+lensing analysis of the data does yields rather weak constraints on $Y_1$, compared to other probes (e.g stellar scales). It should be emphasised that, current upper bounds on $Y_1$ from stellar physics are $\sim \mathcal{O}(0.1)$, however, we choose to quote the actual bound $Y_1 \leq 4.85$ without imposing any prior. Contrary to $Y_1$, the new constraints on the coupling $Y_2$ provide a {\bf 2-fold improvement} compared to current bounds in the literature. The last column indicate the $\Delta\chi^2$ between the cases analysed and the reference model.}
\end{center}

\end{table*}

\begin{figure}
\centering
\includegraphics[width=\columnwidth]{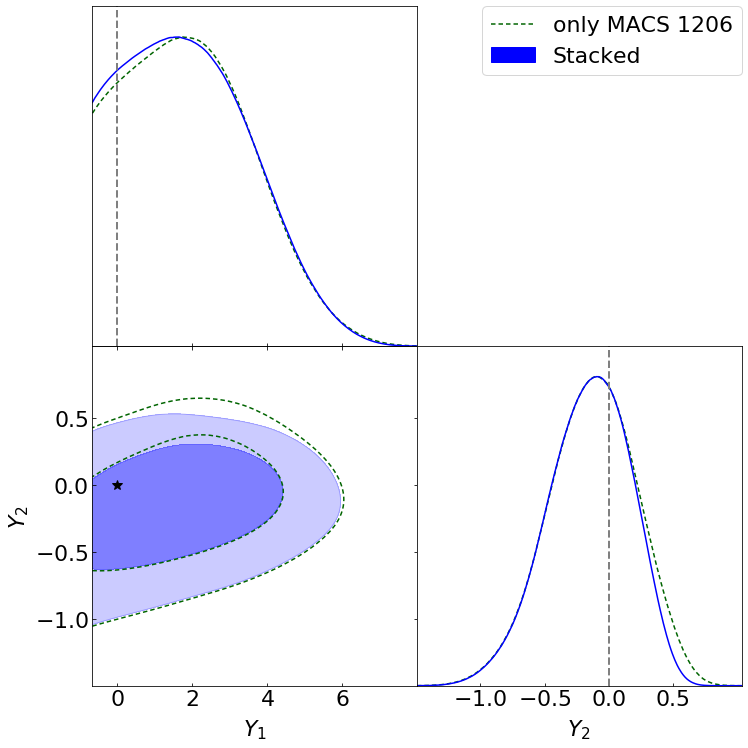}
\caption{\label{fig:stack} Blue lines and shaded areas: marginalized distributions of the two coupling constant $Y_1$ and $Y_2$ obtained from the from the joint lensing+kinematic analysis of MACS 1206 combined with the additional information from the 15-cluster stacked lensing profile. Green-dashed lines: joint lensing+kinematic analysis of MACS 1206 only. The results are shown for the ''ML" anisotropy model. The inner darker region and the outer lighter area refer to $1\, \sigma$ and $2\, \sigma$ regions respectively.}

\end{figure}

\begin{figure}
\centering
\includegraphics[width=\columnwidth]{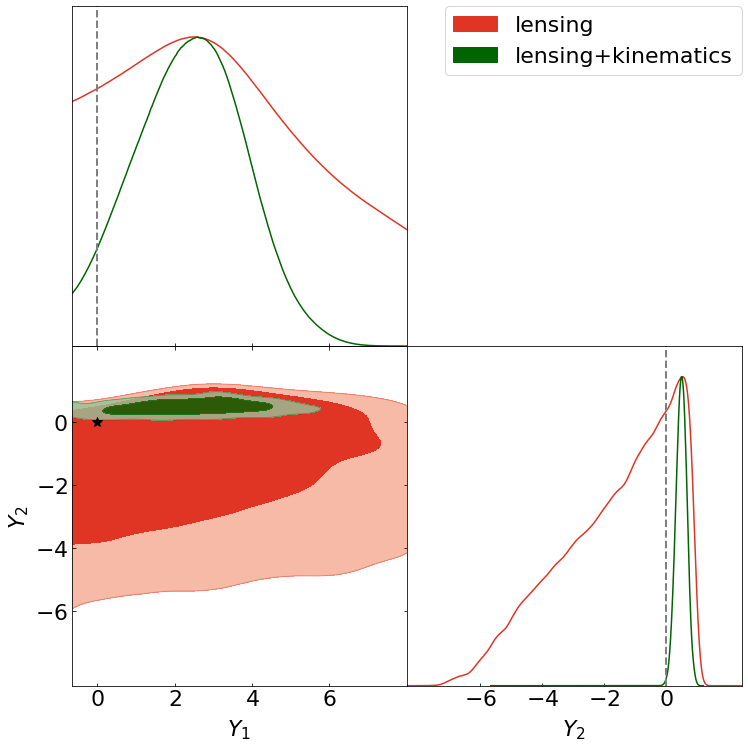}
\caption{\label{fig:resultRXJ} Marginalized distributions of the two coupling constant $Y_1$ and $Y_2$ obtained from the lensing analysis (red) and from the joint lensing+kinematic analysis (green) of Abell~S1063 with a ''O" anisotropy model. The inner darker regions and the outer lighter areas refer to $1\, \sigma$ and $2\, \sigma$ regions respectively. The black star and the vertical black dashed lines indicate the GR expectation values $Y_1=0,\,Y_2=0$. As was the case of MACS1206, constraints on $Y_2$ from the combined kinematics+lensing are competitive, and highly improved compared to those derived based on lensing alone. }

\end{figure}
\subsubsection*{MACS 1206}

In the case of MACS 1206, the results for the modified gravity parameters are in agreement with GR expectations within $1\sigma$, although there is a modest (but statistically irrelevant) preference to positive values for both couplings.
The upper limit of $ Y_1<4.85$ indicates a slight improvement with respect to the internal kinematics analysis only, which is however too weak to be competitive with the current astrophysical bounds. On the other hand, the relativistic coupling is constrained to be $-0.12^{+0.66}_{-0.67}$ (2$\sigma$), which reduces by a factor $\sim 2$ the previous results of \citet{Sakstein:2016ggl}, obtained by a joint lensing and X-ray analyses of the same model over a sample of 58 galaxy clusters. The bounds are also comparable to the more recent constraints of \citet{Laudato21}, which uses lensing and X-ray data of 13 X-ray-selected and 3 lensing-selected clusters from the CLASH sample, including MACS 1206 and Abell~S1063, to fit a generalization of the model discussed in this paper in terms of two parameters $\alpha_H, \,\beta_1$. It is interesting to notice that the constraints obtained by \citet{Laudato21} for MACS 1206 seems to indicate a slight tension with GR which is not found by our analysis.

We further combine our kinematics+lensing chain of MACS 1206 with the posterior obtained from the 15-cluster stacked lensing profile, fitted with the MG-NFW model. In Figure \ref{fig:stack} we show the marginalized distributions of $Y_1$ and $Y_2$ obtained from such analysis (blue lines and blue shaded areas), compared with the single-cluster results. While the posterior of $Y_1$ remains unchanged, a very tiny reduction of the bounds can be found on $Y_2=-0.12^{+0.60}_{-0.63} $. Such a modest improvement is not surprising, as the essential information comes from the combination of lensing and internal kinematics of the same cluster, required to break the degeneracy between the mass profile and the modified gravity parameters. 

\subsubsection*{Abell~S1063}
The joint lensing+kinematic analysis of Abell~S1063 produces an overall $2\sigma$ tension with GR, which is particularly evident in the case of the relativistic coupling. We obtain $Y_1= 2.4^{+2.5}_{-2.8}\,,\;Y_2=0.48\pm 0.34$ at 95\% C.L.; this is a consequence of the fact that internal kinematics analyses suggest a slightly larger $r_{200}$ with respect to lensing (see also the GR mass reconstructions of \citealt{Sartoris20} and \citealt{Umetsu16}), i.e. a "weaker" lensing potential $\Phi_\text{lens}$ compared to $\Phi$. This weakening can be achieved by a positive $Y_2$ in the relativistic potential, which alleviates the discrepancy between lensing and kinematics. 

The tension found by our analysis confirms the results of \citet{,Pizzuti:2017diz} and \citet{Laudato21} which seem to indicate a preference of modified gravity with respect to GR for this cluster. However, is important to point out that Abell~S1063 is not dynamically relaxed at all. The recent analysis of \citet{Mercurio21} found the presence of an asymmetry in the velocity distribution with two peaks at redshift $z_1=0.3413$ and $z_2=0.3555$. Moreover, they identify five sub-clumps with different velocity distribution and a NE-SW elongation in the spatial distribution, supporting the previous findings of \citet{Gomez12}. 
According to \citet{Pizzuti19b}, quantifying deviations form Gaussianity of the l.o.s. velocity distribution through the Anderson-Darling test, can serve as a criterion to select the suitable galaxy clusters on which applying our method. In particular, large values of the Anderson-Darling coefficient $A^2\gtrsim 0.7$ are associated to high probability to find spurious signatures of modified gravity. While in the case of MACS 1206 the Anderson-Darling coefficient is $A^2=0.64$, for Abell~S1063 we found $A^2=1.04$, which is an additional indication of the systematic nature of the tension.
It is nevertheless interesting to note that the higher number of tracers in the \textsc{MG-MAMPOSSt} fit of Abell~S1063 data provides a $\sim 10\%$ reduction of the statistical uncertainties on average, compared to MACS 1206.  

\subsubsection*{Effect of number density and anisotropy modeling}
\begin{figure}
\centering
\includegraphics[width=\columnwidth]{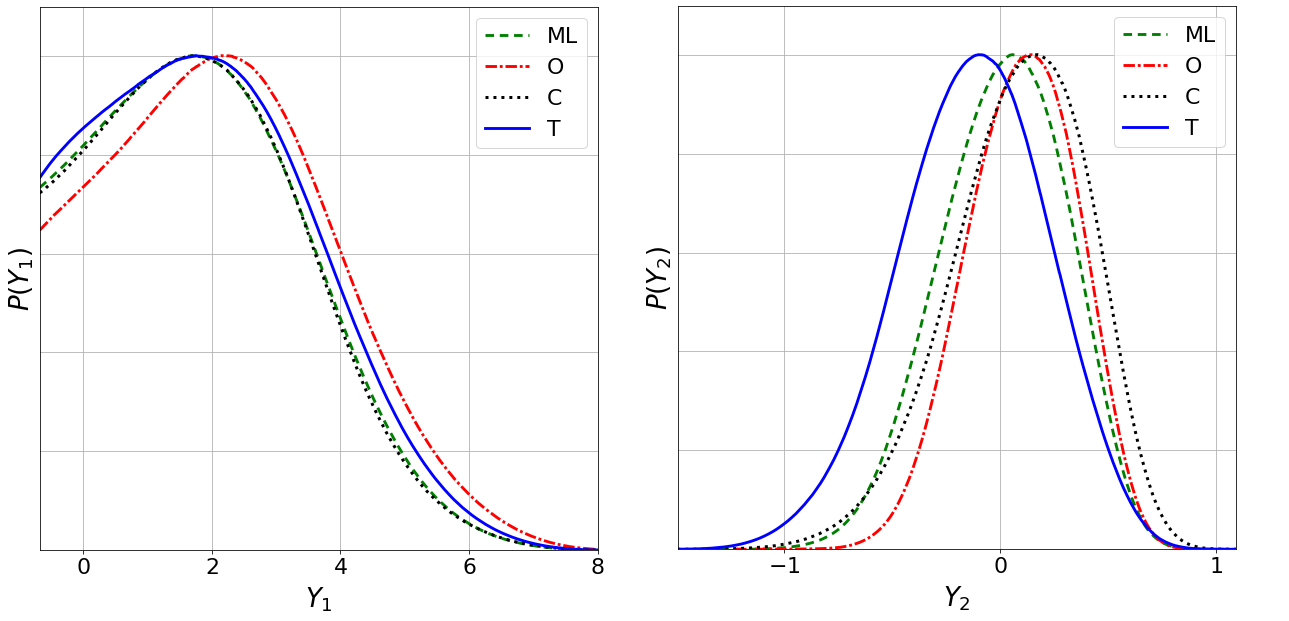}
\caption{\label{fig:MACSallbeta} Marginalized one-dimensional distributions $P(Y_1)$ (left panel) and $P(Y_2)$ (right panel) from the joint lensing+kinematic analysis of MACS 1206 varying the model of the velocity anisotropy profile in the kinematics analysis. Different curves correspond to different anisotropy model with the solid line indicating the reference model. The somewhat larger dependence of the marginalised posterior $P(Y_1)$ on the choice of the anisotropy profile compared to $P(Y_2)$ is of statistical origin and relates to the smaller systematic uncertainties associated with $Y_2$.}

\end{figure}

\begin{figure}
\centering
\includegraphics[width=\columnwidth]{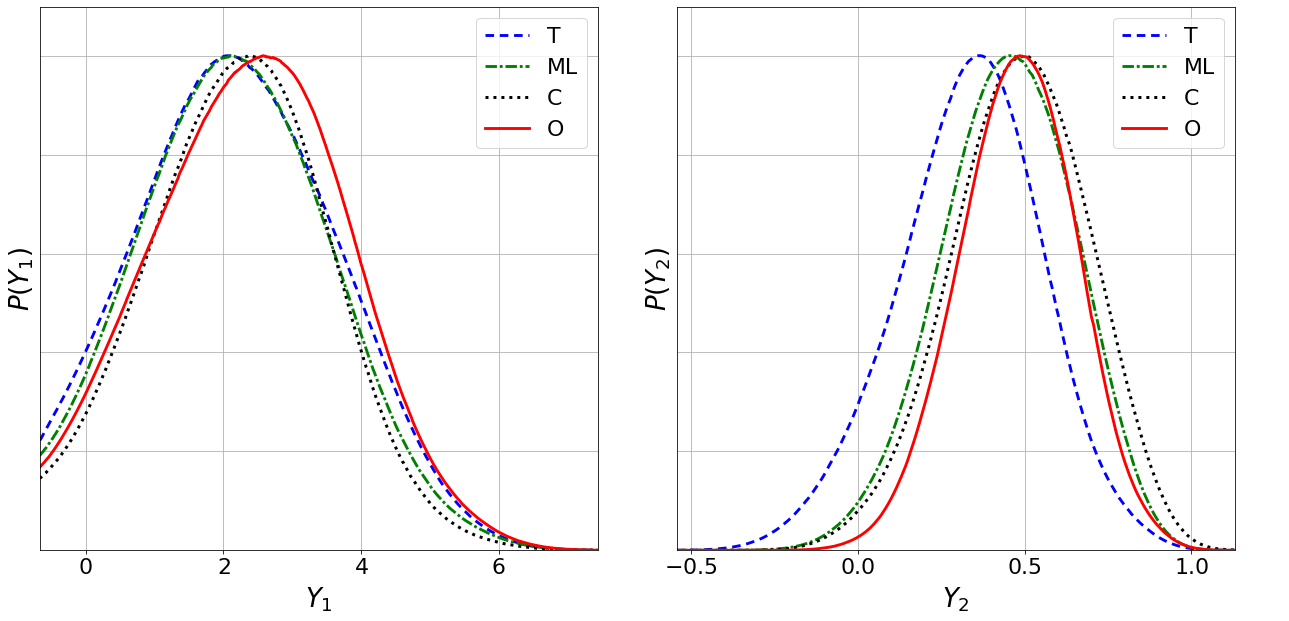}
\caption{\label{fig:RXJallbeta} Marginalized one-dimensional distributions $P(Y_1)$ (left panel) and $P(Y_2)$ (right panel) from the joint lensing+kinematic analysis of Abell~S1063 varying the model of the velocity anisotropy profile in the kinematics analysis, color-coded as in Figure \ref{fig:MACSallbeta}. }

\end{figure}

As a final step, we address the effect of changing the model of the velocity anisotropy profile and the value of the number density profile of the tracers $r_\nu$ within the $1\sigma$ limit. The results are listed in Table \ref{tab:results}, while in Figure \ref{fig:MACSallbeta} (MACS 1206) and Figure \ref{fig:RXJallbeta} (Abell~S1063) we further show the marginalized distributions of $Y_1$ and $Y_2$ for the different $\beta(r)$ prescriptions adopted in this work. Note that the systematics induced by the variation of the anisotropy model are relatively tiny and do not considerably change the outcomes of our analysis. The larger effect has been found for $Y_2$, but this is not surprising as the statistical uncertainties on this parameter are roughly one order of magnitude smaller than $Y_1$ and the role of systematics appear to be more evident.

As for the number density profile, we restrict the analysis to the reference anisotropy model for each cluster and we determine the effects on the marginalized posteriors $P(Y_1)$ and $P(Y_2)$. In this case, given the tight constraints on $r_\nu$, the variation is almost negligible within the statistical uncertainties, as also shown in Table \ref{tab:results}.

As a final result we quote
\begin{equation} \label{eq:constraintsMACS}
Y_1<4.85\,(\text{stat}) \pm 0.2 (\text{syst})\,\;    Y_2=-0.12^{+0.66}_{-0.67}\, (\text{stat})\, \pm 0.21\, (\text{syst}),
\end{equation}
for MACS 1206 and 
\begin{equation} \label{eq:constraintsRXJ}
    Y_1= 2.4^{+2.5}_{-2.8}\,(\text{stat})\pm 0.6 (\text{syst}) \,,\;Y_2=0.48\pm 0.34\, (\text{stat})\,\pm 0.22 \,(\text{syst})
\end{equation}
for Abell~S1063, where the systematic uncertainties are estimated as the absolute value of the largest shift in the peak of the marginalized posteriors.  

\section{Summary and Discussions} \label{sec:summary}
In this paper we have used high-quality mass profile determinations of galaxy clusters inferred from kinematics of member galaxies and the combination of strong and weak lensing, for the CLASH clusters MACS 1206 and Abell~S1063, to constrain the lensing potential within the general scalar-tensor theories for dark energy, known as DHOST, characterized by a partial breaking of the Vainsthein screening mechanism. This work is an extension of our forecasting analysis presented in \citet{Pizzuti2021} based on the recently developed code \textsc{MG-MAMPOSSt}. In the case of MACS 1206, we have further combined the kinematic+lensing posterior with the stacked lensing profile of a CLASH X-ray-selected subsample of 15 clusters, as external information when statistically inferring the values of the fifth force parameters (Figure~\ref{fig:stack}).

Within the family of theories that we studied here, the new scalar degrees of freedom affect both the kinematics and lensing of galaxy clusters through a modification of the relevant Poisson equations. Our analysis shows that, the combination of lensing and kinematics provides competitive new constraints on the strength of the lensing potential in these theories, compared to the lensing observations alone. The new constraints on the dimensionless coupling $Y_2$, which controls the modification in the lensing potential (see equations (\ref{eq:BH_Mamon}) and (\ref{eq:BH_lensing})) are summarised in eq. (\ref{eq:constraintsMACS}) and eq. (\ref{eq:constraintsRXJ}). These results take into account the uncertainties due to the parametrization of the velocity anisotropy profile and to the value of the scale radius in the number density profiles of galaxies in the kinematic analysis. For MACS 1206, the constraints are in agreement with GR expectation,  providing about a 2-fold improvement of previous analyses based on the combination of gas dynamics and weak lensing of galaxy clusters \citet{Sakstein:2016ggl}. From Abell~S1063 a $>2\sigma$ tension with GR arises, independently of the anisotropy model chosen in the analysis. This is probably a spurious effect due to the un-relaxed state of the cluster which has been suggested by previous analyses (e.g \citealt{Mercurio21}), further confirmed by the application of the selection criterion of \citet{Pizzuti19b} which indicates that Abell~S1063 is not a good candidate for the application of the proposed method.

Our work can be extended in different ways. First, the kinematics analysis can be applied to a larger set of galaxy clusters, further including the total mass profile as obtained from the pressure and temperature gas profiles extracted from X-ray data. Both hot diffuse gas and galaxies in clusters feel the time-time gravitational potential $\Phi$, but they are affected by different physical processes. 

The combination of lensing, member galaxy kinematics and X-ray analyses may help in obtaining tighter constraints on the MG parameters and in better exploring how the assumption of dynamical relaxation affects the  final results. This will be fundamental in view of the upcoming sky surveys, performed thanks to the next-generation instruments mounted on space (e.g. Euclid, JWST) and ground based (e.g.  Vera C. Rubin Observatory) telescopes, which will provide a large amount of data to reconstruct mass profiles of several galaxy clusters.

Moreover, as already discussed in \citet{Pizzuti2021}, the NFW model may not be the best representation of the total matter density profiles of clusters in modified gravity; other models could offer a more realistic description of the behavior of the gravitational interaction at cluster's scales. In particular, the Vainsthein screening can be less or more efficient depending on the shape of the matter distribution (e.g. \citealt{Burrage19}).
Furthermore, it would be interesting to investigate the redshift dependence of the fifth force couplings by studying a sample of galaxies in different redshift bins. We leave this issues for future work.

\bibliographystyle{mnras}
\bibliography{master}


\section*{Acknowledgements}
LP is partially supported by a 2019 "Research and Education" grant from Fondazione CRT. The OAVdA is managed by the Fondazione Cle\'ment Fillietroz-ONLUS, which is supported by the Regional Government of the Aosta Valley, the Town Municipality of Nus and the "Unite\' des Communes valdotaines Mont-E\'milius.
I. D. Saltas is supported by the Grant Agency of the Czech Republic (GAČR), under the grant number 21-16583M. K.U. acknowledges support from the Ministry of Science and Technology of Taiwan (grant MOST 109-2112-M-001-018-MY3) and from the Academia
Sinica (grant AS-IA-107-M01). 
This project was partially funded by PRIN-MIUR 2017 WSCC32.

\section*{Data Availability}

The data underlying this article were provided in part by \citet{Mercurio21,Sartoris20,Umetsu16,Biviano01} by permission. Data will
be shared on reasonable request to the corresponding author
with the permission of CLASH-VLT collaboration: \citet{Rosati2014,Umetsu16,Mercurio21}.


\bsp	
\label{lastpage}
\end{document}